\documentclass[fleqn,10pt,twocolumn]{wlscirep}
\usepackage[utf8]{inputenc}
\usepackage[T1]{fontenc}

\newcommand{\doi}[1]{\href{https://doi.org/#1}{#1}}

\let\Re\undefined

\DeclareMathOperator{\Re}{Re}
\DeclareMathOperator{\Tr}{Tr}

\graphicspath{{./arxiv_figs/}}

\title{Advances in machine-learning-based sampling motivated by lattice quantum chromodynamics}

\author[1]{Kyle Cranmer}
\author[2]{Gurtej Kanwar}
\author[3]{S\'ebastien Racani\`ere}
\author[3]{Danilo J. Rezende}
\author[4,5,*]{Phiala E. Shanahan}
\affil[1]{Physics Department, University of Wisconsin-Madison, Madison, WI, USA}
\affil[2]{Albert Einstein Center for Fundamental Physics, Institute for Theoretical Physics, University of Bern, Bern, Switzerland}
\affil[3]{Google DeepMind, London, UK}
\affil[4]{Center for Theoretical Physics, Massachusetts Institute of Technology, Cambridge, MA, USA}
\affil[5]{The NSF AI Institute for Artificial Intelligence and Fundamental Interactions, Cambridge, MA, USA}

\affil[*]{e-mail: pshana@mit.edu}

\begin{abstract}
Sampling from known probability distributions is a ubiquitous task in computational science, underlying calculations in domains from linguistics to biology and physics. 
Generative machine-learning (ML) models have emerged as a promising tool in this space, building on the success of this approach in
applications such as image, text, and audio generation. Often, however, generative tasks in scientific domains have unique structures and features---such as complex symmetries and the requirement of exactness guarantees---that present both challenges and opportunities for ML.
This Perspective outlines the advances in ML-based sampling motivated by lattice quantum field theory, in particular for the theory of quantum chromodynamics. Enabling calculations of the structure and interactions of matter from our most fundamental understanding of particle physics, lattice quantum chromodynamics is one of the main consumers of open-science supercomputing worldwide. The design of ML algorithms for this application faces profound challenges, including the necessity of scaling custom ML architectures to the largest supercomputers, but also promises immense benefits, and is spurring a wave of development in ML-based sampling more broadly. In lattice field theory, if this approach can realize its early promise it will be a transformative step towards first-principles physics calculations in particle, nuclear and condensed matter physics that are intractable with traditional approaches.

\end{abstract}

\begin{document}

\flushbottom
\maketitle

\thispagestyle{empty}

\section{Introduction}

Theoretical nuclear physics has the ironic feature that although the fundamental laws are well understood, the computations required to make quantitative, first-principles predictions are in many cases currently infeasible. The strong nuclear force is fundamentally described by the quantum field theory known as Quantum Chromodynamics (QCD), which details the dynamics of constituent particles---quarks and gluons---that arise as excitations of underlying quantum fields. This theory successfully predicts a wide range of phenomena that occur at different energy scales, ranging from the high-energy collisions at the Large Hadron Collider to the properties and interactions of composite particles such as the proton and neutron, as well as the nuclei they form. At high energies, the interactions between quarks and gluons are weak, and accurate QCD calculations can be made using a perturbative expansion, which is often represented with Feynman diagrams. At the lower energies relevant for much of nuclear physics, the interactions between quarks and gluons are strong and the perturbative approach breaks down. In this regime, quantitative predictions can be achieved through a computational approach known as lattice QCD, in which the quark and gluon fields are represented on a discrete spacetime lattice.
Many key aspects of nuclear physics can be computed precisely in this framework. For example, such calculations reveal how the masses of the proton and neutron arise from the fundamental quarks and gluons~\cite{Borsanyi:2014jba}, and they have been used to make predictions of the masses of new composite particles later discovered by experiments at CERN~\cite{Brown:2014ena,LHCb:2014nae,LHCb:2017iph}. However, the reach of this approach is limited by its computational cost, and controlled first-principles QCD calculations of nuclear structure and reactions, for example, would require a scale of computational resources that is currently infeasible~\cite{Joo:2019byq}.
Without breakthrough developments, many important studies will remain impossible even with the world's next generation of exascale supercomputers (quintillions ($10^{18}$) of operations per second, or the equivalent of 50 million laptops working in concert). If the computational cost of lattice field theory can be greatly reduced, fundamental questions in particle, nuclear and condensed matter physics will be answered. For example, first-principles calculations can probe the fine-tunings in nuclear physics that are deeply important for understanding our existence, by revealing how sensitive the production of carbon in the Universe via the triple-$\alpha$ process is to the free parameters of the theory, explaining why protons and neutrons cluster inside nuclei, and elucidating how the lightest elements formed in the first minutes of the Universe’s existence via Big Bang nucleosynthesis~\cite{Detmold:2019ghl}. 

Calculations in lattice QCD are cast in the form of statistical averages with respect to a distribution of quark and gluon field configurations. A major component of the computational cost of lattice QCD calculations is the estimation of these averages by Monte Carlo sampling techniques. (Sampling is one of several computationally-intensive steps in lattice QCD calculations. Others, such as the inversion of Dirac operators for the calculation of physical observables, may also be accelerated using machine learning (ML) approaches~\cite{Cali:2022qbd,Lehner:2023prf,Lehner:2023bba}). Sampling representative configurations of a system to quantitatively evaluate its properties is ubiquitous in physics, being used in fields spanning from ab-initio molecular dynamics and statistical physics to astrophysics, and many others. However, sampling from the highly-structured, high-dimensional, and multi-modal distribution of configurations in lattice QCD presents an extraordinarily difficult computational challenge. 
This problem has historically been the impetus for the development of what have become foundational techniques in computational statistics and high-performance computing, with far-reaching implications within and beyond physics. For example, both the classic Metropolis--Hastings Markov chain Monte Carlo algorithm~\cite{Metropolis:1953am} and Hamiltonian/hybrid Monte Carlo (HMC)~\cite{Duane:1987de} were first developed in the context of theoretical nuclear physics, with the latter conceived specifically for lattice QCD.
Similarly, the IBM Blue Gene series of supercomputers trace their origins back to the QCDOC (quantum chromodynamics on a chip) computer built specifically for this particular application~\cite{Chen:2000bu}.

The rapid advance of ML over the past few years has spurred the emergence of a new class of algorithms that are revolutionizing computing for both science and industry applications. For example, the extraordinary success of the ML tool AlphaFold~\cite{jumper2021highly} in protein folding took the world of biology by surprise, redefining the pace of progress in a field where algorithmic developments had been slow for decades.
For lattice QCD, which has historically driven a virtuous cycle of innovations in scientific computing, these advances promise a new chapter. In particular, the rise of generative modelling with ML~\cite{hoffmann2022training,thoppilan2022lamda} suggests the particular application of sampling algorithms for lattice QCD.
The sampling problem in lattice QCD has several key features that present both challenges and opportunities to ML. On the one hand, any algorithm must be asymptotically exact, preventing the direct application of certain generative ML approaches such as generative adversarial networks or variational autoencoders (VAEs). A practical challenge is also presented by the extreme scale of lattice QCD samples used in state-of-the-art calculations, each of the order of several terabytes at the current time. On the other hand, the forms of the relevant probability distributions are exactly known, which can inform the design and training of sampling architectures. In particular, these distributions are invariant under complicated and high-dimensional symmetry groups which significantly reduce the dimensionality and complexity of the problem if they can be incorporated exactly. Although it has required considerable effort to develop ML models that incorporate the symmetries of lattice QCD into ML architectures, the investment has paid dividends in the efficacy of the resulting algorithms.

This Perspective reviews the unique requirements and features of a class of ML-based sampling strategies that have been recently developed for lattice QCD applications and places these developments in the broader context of ML for sampling in scientific domains. Although this endeavour remains in its early stages, it is already clear that it has considerable potential, not only to emulate the transformative impact that ML has had in applications such as AlphaFold~\cite{jumper2021highly}, but also to spur the advancement of ML itself. 

\section{Lattice QCD and the sampling problem}\label{sec:sampling_in_lqcd}

\newcommand{\qfield}[0]{\Phi} 

The lattice method for computing physical observables in quantum field theories such as QCD proceeds by discretizing space and time onto a four-dimensional grid (or `lattice'), with spacing $a$ between neighbouring points and a finite volume $V$. 
In this framework, the fundamental particle degrees of freedom of the theory--- quarks and gluons in QCD---are represented through `quantum fields' that consist of complex numbers, vectors or matrices associated with the points and edges (or `links') of the lattice. Quantities of physical interest are then defined by integrals over these field degrees of freedom, and
the continuum, infinite-volume theory is recovered by taking the limit $a\rightarrow 0$, $V\rightarrow \infty$.

A general physical observable can be defined in terms of quantum `operators' $\mathcal{O}$ and computed as a statistical expectation value\cite{Peskin:1995ev:EuclPI}:
\begin{align} \label{eq:expectval}
    \langle \mathcal{O}\rangle  = \int \mathcal{D}\qfield \, \mathcal{O}[\qfield]p[\qfield],\text{  where } p[\qfield] = e^{-S[\qfield]}/Z.
\end{align}
Here the notation $\int \mathcal{D}\qfield$ schematically indicates integration over all configurations of the discretized quantum fields collectively denoted by $\qfield$, and $Z = \int \mathcal{D}\qfield e^{-S[\qfield]}$ is a normalizing constant. The `action' $S[\qfield]$ encodes the dynamics of the theory by defining the statistical distribution $p[\qfield]$; in QCD, it describes the fluctuations and interactions of the quark and gluon fields. The operator $\mathcal{O}$ can be chosen to study various physical properties of the theory; for example, the mass of the proton can be calculated using an operator that represents the interaction of two up quarks and one down quark.

In practice, the integral in equation~\eqref{eq:expectval} cannot be computed analytically and is instead evaluated by Monte Carlo integration, that is, using an ensemble of $N$ field configurations $\{\qfield_1, \dots, \qfield_N\}$ sampled from the distribution $p[\qfield]$. Physical quantities are then computed as $\langle \mathcal{O}\rangle \approx  \frac{1}{N}\sum_{i=1}^N\mathcal{O}[\qfield_i]$ with an uncertainty that is systematically improvable by taking $N$ large. The first step of any lattice field theory calculation is thus a sampling problem. Although the challenge of generating lattice field configurations is reminiscent of sampling problems in many other fields, the structure of the quantum fields, the complicated symmetries of the distribution $p[\qfield]$ and the sheer scale of the required calculations set this apart as a particularly difficult computational problem.

\subsection{Structure and symmetries of field configurations}

\begin{figure}[ht!]
	\centering
	\includegraphics[width=0.95\linewidth]{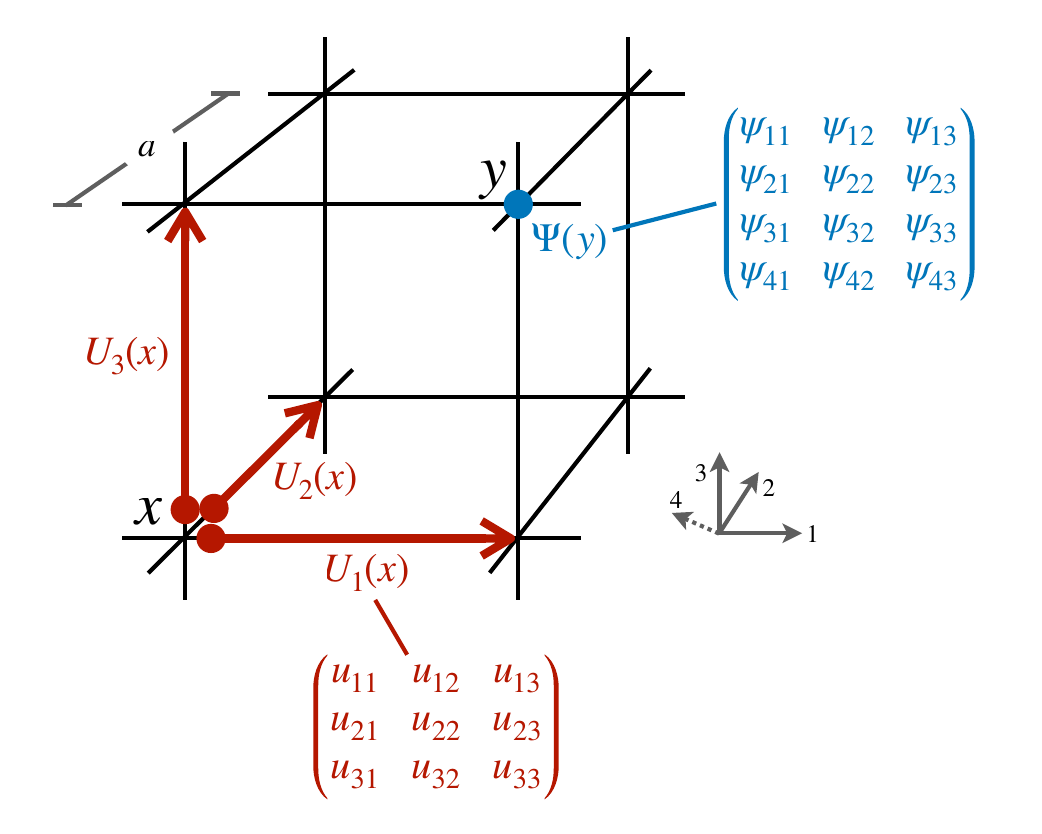}
	\caption{
	    \textbf{Depiction of a single cube within the spacetime lattice of a lattice QCD calculation.} Shown are some elements $U_\mu(x)$ of the discretized gluon field (red), each associated with an edge $(x,x+\hat{\mu})$ from site $x$ to the neighboring site in direction $\mu \in \{1,2,3,4\}$, and an element $\Psi(y)$ of the discretized quark field (blue), associated with a site $y$. The value of each $U_\mu(x)$ is a complex unitary $3 \times 3$ matrix with determinant $1$, that is, an SU(3) matrix, and each $\Psi(y)$ is a $4 \times 3$ complex matrix. $a$ is the lattice spacing between neighbouring points. The fourth dimension of the lattice is suppressed for clarity.
	}
	\label{fig.lattice}
\end{figure}

In typical lattice quantum field theories, the discretized quantum fields not only extend over the spacetime lattice, but also have `internal' degrees of freedom represented mathematically by a vector or matrix structure at each point or edge of the lattice. In particular, in QCD the gluon field $U$ is encoded by $SU(3)$ variables---$3 \times 3$ complex unitary, unit-determinant matrices---on each edge of the lattice, whereas the quark fields $\Psi$ are encoded by $4 \times 3$ complex matrices on each site of the lattice, as shown in Fig.~\ref{fig.lattice}. 
For QCD, the calculation of a physical observable via equation~\eqref{eq:expectval} can thus be expressed as
\begin{equation} \label{eq:expectval-2}
\begin{gathered}
\begin{aligned}
    \langle \mathcal{O}\rangle &= \frac{1}{Z}\int \mathcal{D}U \mathcal{D}\bar{\Psi} \mathcal{D}\Psi \mathcal{O}[U,\bar{\Psi},\Psi]e^{-S[U,\bar{\Psi},\Psi]} \\
    &= \frac{1}{Z} \int \mathcal{D}U \mathcal{O}'[U] e^{-S_{\mathrm{eff}}[U]},
\end{aligned} \\
    \text{where }
    Z = \int \mathcal{D}U \mathcal{D}\bar{\Psi} \mathcal{D}\Psi e^{-S[U,\bar{\Psi},\Psi]}.
\end{gathered}
\end{equation}

Here the notation $\int \mathcal{D}U$ indicates integration over all values of the discretised gluon field $U$, whereas the integration $\int \mathcal{D}\bar{\Psi} \mathcal{D}\Psi$ over all values of the discretized quark fields are Gaussian integrals that are evaluated analytically, yielding a modified operator $\mathcal{O}'$ and the modified weight $p[U]=e^{-S_{\mathrm{eff}}[U]}/Z$ over gluon field configurations. (In particular, the integral $\int \mathcal{D}\bar{\Psi} \mathcal{D}\Psi$ is a Berezin integral\cite{Berezin:1966nc} over elements of a Grassmann algebra, which must be analytically treated to produce an integral amenable to numerical evaluation.)
In practice, auxiliary degrees of freedom known as `pseudo-fermions'~\cite{Gattringer:2010zz} are also typically introduced as stochastic estimators for determinants appearing in $p[U] = \exp(-S_{\mathrm{eff}}[U])/Z$.
State-of-the-art lattice QCD calculations involve fields of size up to $256^3 \times 512 \approx 8.6 \text{ billion}$ lattice sites
with quantum fields represented by roughly $50$ degrees of freedom per lattice site (this counting includes four $SU(3)$ matrices for each lattice site, yielding $4\times 8=32$ degrees of freedom, as well as  complex $4\times 3$ matrices with $2\times 3\times 4=24$ degrees of freedom for each site, arising from the pseudo-fermion fields), meaning that, in practice, calculations involve Monte Carlo integration over as many as $10^{12}$ variables.

Symmetries in a lattice field theory manifest as transformations of field configurations that leave the probability density $p[U]$ and the integration measure invariant.
The action, and hence $p$, is typically invariant under both discrete geometric symmetries of the hybercubic Euclidean spacetime, such as discrete translations, rotations and reflections, and under internal symmetry transformations. For example, one contribution to the lattice QCD action is given by
\begin{equation}
    S_g[U] = -\frac{\beta}{6} \sum_x \sum_{\substack{\mu = 1 \\  \nu  = 1}}^{4} \Re \Tr [U_\mu(x) U_\nu(x + \hat{\mu}) U^\dag_\mu(x + \hat{\nu}) U^\dag_\nu(x)],\label{eq.pure_gauge_action}
\end{equation}
where $\beta$ is a parameter of the theory that is related to the lattice spacing $a$, $x$ is summed over the sites of the discretized lattice, and $\hat{\mu},\hat{\nu}$ indicate vectors of length $a$ in the $\mu$ and $\nu$ directions, respectively (see Fig.~\ref{fig.lattice}). From this expression, it can be seen how `gauge' symmetry is manifest in QCD, as $p[U]$ is invariant under the transformation of the gauge field $U$ according to 
\begin{equation} \label{eq:gauge-transform}
    U_\mu(x) \rightarrow \Omega(x) U_\mu(x) \Omega^\dagger(x+\hat{\mu})
  \end{equation}
for all possible choices of $\Omega(x) \in \text{SU(3)}$ over all lattice sites. Because this symmetry is specified by one SU(3)-valued matrix per lattice site (so eight degrees of freedom per site), the symmetry group may have a dimension as large as $10^{11}$ in state-of-the-art calculations.

\subsection{Approaches and challenges to sampling field configurations}
Conventionally, the generation of an ensemble of lattice fields distributed according to $p[\qfield]$ is performed iteratively using a Markov process, in which a chain of configurations $\{\qfield_1, \qfield_2,\ldots\}$ is generated by a sequence of stochastic updates beginning from an initial configuration $\qfield_0$. In particular, the HMC algorithm was first conceived of in the 1980s specifically for this application in lattice field theory~\cite{Duane:1987de} and has since become a mainstay of the computational science community. In this paradigm, the rapid exploration of the state space is achieved by a directed evolution from each configuration to a new proposed configuration, which avoids an inefficient random walk. Exactness of the distribution is guaranteed by applying the Metropolis--Hastings procedure to accept the proposed configuration with an appropriate probability~\cite{Metropolis:1953am,hastings1970monte} (see also the next section).

Despite the outstanding success of this approach --- which remains the workhorse of lattice field theory --- generating ensembles of field configurations is one of the notable computational costs of first-principles QCD calculations. In particular, because the approach evolves configurations via
a local dynamical process,
increasingly many updates are required to decorrelate samples on physical length scales as the continuum limit is approached ($a\rightarrow 0$). This is a manifestation of the phenomenon known as `critical slowing-down' in this context~\cite{Duane:1987de,Schaefer:2010hu}. Simultaneously, the distribution of QCD gauge fields spans topologically distinct sectors, and Markov-based sampling algorithms such as HMC can become `trapped' or `frozen' in sectors of fixed topology.

Any alternative approach to sampling lattice field configurations will need to satisfy several key requirements in order to be practically viable. Most importantly, it must be statistically improvable: that is, the true probability distribution $p[\qfield]$, including the various symmetries that this distribution respects, must be recovered in the limit of a large number of samples. Furthermore, the approach must be efficiently scalable to state-of-the-art lattice field theory studies, which involve field configurations as large as many terabytes of memory each, with as many as $10^{12}$ degrees of freedom. Finally, the approach must improve upon the impressive success of the HMC framework and mitigate the challenges of critical slowing-down and topological freezing in some regime of physical interest.

\begin{figure*}[t]
	\centering
	\includegraphics[width=0.95\textwidth]{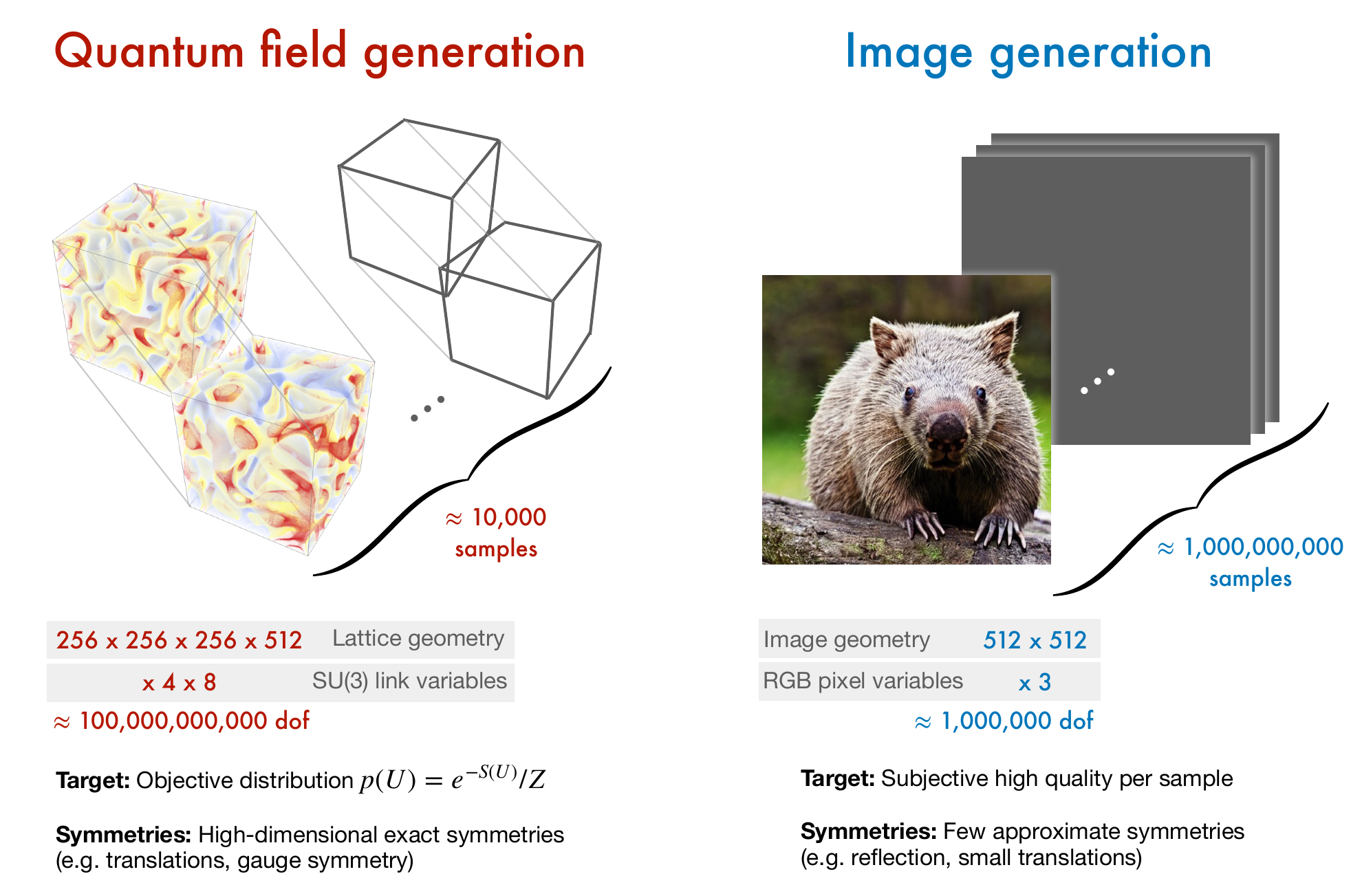}
	\caption{\textbf{Comparison between the sampling tasks of quantum field generation for lattice quantum chromodynamics and image generation.} In addition to differences in the target and symmetries of the problems, the hierarchy of degrees of freedom (dof) per sample to number of samples is inverted for quantum field generation as compared with image generation.
	The action $S$ encodes the dynamics of the theory by defining the statistical distribution $p$, $U$ the gluon field, and $Z$ is a normalizing constant.
    The image on the right side is reprinted from Kaggle (\href{https://www.kaggle.com/datasets/vitaliykinakh/stable-imagenet1k}{https://www.kaggle.com/datasets/vitaliykinakh/stable-imagenet1k}) under a Creative Commons license (\href{https://creativecommons.org/publicdomain/zero/1.0/}{CC0 1.0}).}
	\label{fig:ml_image_vs_lattice}
\end{figure*}

\section{ML for sampling lattice field configurations}\label{sec:ml_for_sampling}

An ML approach to sampling lattice field configurations is an appealing proposition: it offers a new paradigm of algorithms that are optimised specifically for the task at hand. Enabling radically different approaches to sampling, ML may mitigate critical slowing-down and other key challenges faced by traditional Markov-process algorithms such as HMC. Even for cases where HMC works well, ML may still provide advantages, for example by enabling embarrassingly parallel, rather than sequential, sampling, or by learning to approximate a large number of computational steps of traditional algorithms with fewer operations, as is observed in other fields such as ML approximations of partial differential equations solvers\cite{beck2020overview}.

However, the application of ML to lattice field theory is not straightforward, given the previously highlighted features of the lattice field theory problem.
In particular, for the theory of QCD, gauge field samples are collections of matrices that are constrained to be SU(3) matrices (see Fig.~\ref{fig.lattice}), whereas samples in typical ML domains such as images or natural language models are represented by vectors of unconstrained real numbers.
When considering generative models based on diffeomorphisms, as discussed below, it is imperative to ensure that these constraints are satisfied by all transformations.
The associated SU(3) gauge symmetry is also very atypical in comparison with usual ML applications, although other symmetries such as 4D translations may be handled by traditional ML methods such as convolutional neural networks.
Moreover, generative models for traditional applications such as images and language do not require asymptotic guarantees of exactness in sampling, whereas these are critical in the lattice field theory context. Finally, the sheer scale of state-of-the-art lattice QCD calculations, both in terms of the scale of lattice samples and the computational cost required to manipulate them, presents a challenge to ML approaches. Figure~\ref{fig:ml_image_vs_lattice} illustrates these stark contrasts between the lattice field generation problem and other sampling tasks that have been revolutionized through ML, such as image generation.  Clearly, achieving state-of-the-art sampling performance with new ML algorithms in the context of lattice field theory will require the development of new algorithms and innovation in ML.

\subsection{Classes of generative models for sampling in lattice field theory}
ML models designed to (approximately) sample from a target density are known as generative or probabilistic models. A generative model typically consists of three components: a space of latent or hidden variables equipped with a density, a set of observed or target variables, and a parametric map that transforms points in the latent space into points in the target space. Optimization is performed on the parametric map so that the density it induces in the target space approximates the target density. A wide variety of ML-based generative architectures have been developed over the past decade, with transformative successes particularly evident in applications to sound/image data \cite{oord2016wavenet, dhariwal2021diffusion,saharia2022photorealistic,child2020very} and language data \cite{hoffmann2022training,kaplan2020scaling,brown2020language,lieber2021jurassic,rae2021scaling,smith2022using,thoppilan2022lamda}.

One notable difference between these applications and the challenge of sampling field configurations for lattice QCD is that the true distribution over the space of images, sounds or text is not known, so the model distribution is learnt from data samples. For lattice field configurations, not only is the unnormalized target distribution known, but it must be sampled from with asymptotic guarantees of exactness. This can be achieved with ML models if they feature tractable likelihoods (the model probability density can be computed for any given sample); in this case, they can be embedded inside sampling algorithms with asymptotic guarantees, such as a Markov chain, as discussed further below.

A tractable likelihood also allows one to optimize an ML model by minimizing a probability divergence $D(q_{\theta}; p)$ between the model probability density $q_{\theta}[U]$ parameterized by $\theta$ and the known target probability density $p[U] = e^{-S_{\mathrm{eff}}[U]}/Z$. However, in stark contrast to typical ML sampling problems, training models for lattice QCD sampling requires estimating the gradients $\nabla_{\theta}D(q_{\theta}; p)$ using only samples from the model or perhaps only a small number of `ground truth' data samples. This restricts the family of probability divergences that can be used; $f$-divergences such as the Kullback--Leibler divergence~\cite{Kullback:1951} are commonly used, as they can be expressed as an expectation value under $q_{\theta}$, allowing the divergence and its gradient to be estimated from model samples alone.

Any ML approach to sampling for lattice QCD must also be `scalable' as the number of lattice sites, $M = V/a^4$, is increased. Ideally, its computational and memory costs should scale linearly or sub-linearly with the number of lattice sites, which we denote by $O(M)$ below. This applies to all aspects of the model: drawing samples, evaluating the likelihood $q_{\theta}$, and evaluating the gradients $\nabla_{\theta}D(q_{\theta}; p)$. This consideration restricts or rules out certain classes of models, as discussed below.

The features of various generative modelling frameworks that could be considered for the lattice QCD sampling problem are outlined below. 

\begin{itemize}
\item {\bf Latent-variable models} such as generative adversarial networks\cite{goodfellow2014generative} and VAEs\cite{rezende2014stochastic,kingma2013auto} typically have efficient $O(M)$ sampling, but intractable likelihoods (involving marginalization of latent variables).

\item {\bf Auto-regressive models} \cite{oord2016wavenet,van2016pixel,hoffmann2022training,kaplan2020scaling,brown2020language,lieber2021jurassic,rae2021scaling,smith2022using,thoppilan2022lamda} typically have efficient $O(M)$ likelihood evaluation. Sampling can also be achieved with $O(M)$ cost in principle, but existing implementations are impractically slow.

\item {\bf Continuous time models} include diffusion models~\cite{dhariwal2021diffusion,saharia2022photorealistic} defined via stochastic differential equations and continuous-time normalizing flows\cite{chen2018continuous} defined via ordinary differential equations. In these models, likelihood computation requires integrating a scalar ordinary differential equation defined by the divergence of a vector field (the marginal score function). This computation typically has computational cost $O(M^3)$ unless additional structure is forced on to the model\cite{chen2019neural}.

\item {\bf Discrete time normalizing flow models}\cite{papamakarios2021normalizing,rezende2015variational,tabak2013family,dinh2016density,kingma2018glow,papamakarios2017masked} remain as good candidate models. In a discrete flow, the generative process maps a latent vector $z$ (a field configuration in the lattice field theory context) sampled from a base density into the target density via the composition of a series of parametric diffeomorphisms $F_1$, $F_2$, \dots, $F_n$. If $z$ is sampled with density $r(z)$, then a flow sample $x = F(z)$ has known density $q(x) = r(z) \left|\det{\partial F / \partial z}\right|^{-1}$, where $F=F_1\circ \dots \circ F_n$ is the composed diffeomorphism. By restricting to $F_i$ for which $\partial F_i / \partial z$ is a triangular matrix, the cost of $\left|\det{\partial F / \partial z}\right|$ is only $O(M)$ (ref.~\citen{papamakarios2021normalizing}).
\end{itemize}

These models, however, also have intrinsic limitations that must be worked around, such as topology preservation
of the diffeomorphism~\cite{huang2020augmented} and difficulty in modelling tail-behaviour of a target density if the tails are not already in the base density~\cite{laszkiewicz2022marginal}.

One can also consider sampling in an augmented space, where the data space is augmented with an additional set of auxiliary or latent variables. In this setting, it may be viable to reconsider VAEs~\cite{rezende2014stochastic,kingma2013auto} or VAE--flow hybrids~\cite{wu2020stochastic}, but currently there are no results demonstrating these methods perform well compared to models working directly on the data space for lattice QCD.

\subsection{Methods to guarantee asymptotic exactness} \label{sec:exact}
Several mechanisms have been proposed to combine generative models with Markov chain Monte Carlo and importance-sampling algorithms in order to inherit their asymptotic convergence guarantees.
One of the simplest mechanisms is neural importance sampling, in which a model density $q_{\theta}(x) \approx p(x)$ is used to evaluate expectations under the target $p$ via $\mathbb{E}_p[\mathcal{O}(x)] = \mathbb{E}_{q_{\theta}}[\frac{p(x)}{q_{\theta}(x)}\mathcal{O}(x)]$ (ref.~\citen{muller2019neural}).
An appealing alternative is to incorporate generative models into an asymptotically exact Markov process, which allows existing analysis techniques to be used or existing Markov chain updates to be combined with the ML sampling approach.
Generally, the Metropolis--Hastings algorithm uses an ergodic transition kernel $K(x'|x)$ to propose Markov chain updates $x \rightarrow x'$ which are accepted with probability 
\begin{equation}
    p_{\mathrm{acc}}(x'|x) = \min\left(1, \frac{K(x|x') p(x')}{K(x'|x) p(x)}\right)
\end{equation}
to ensure that the asymptotic equilibrium distribution is the desired target distribution~\cite{Metropolis:1953am,hastings1970monte}.
Although this method is guaranteed to converge~\cite{robert1999monte}, the speed of convergence depends on the target density and the choice of transition kernel $K(x'|x)$.

ML models can be combined with the Metropolis--Hastings approach by using generative models to construct the kernel $K(x'|x)$. A direct approach is to use the trained model $q_{\theta}(x) \approx p(x)$ to produce independent and identically distributed (iid) proposal samples, $K(x'|x) = q_{\theta}(x')$, with the convergence rate determined by the quality of the model approximation to the target distribution. More advanced techniques include neural transport Monte Carlo\cite{hoffman2019neutra,nijkamp2020learning}, where Markov chain Monte Carlo is performed in the latent space; learned Monte Carlo proposals\cite{wang2018meta,song2017nice,li2021neural,Huang:2017,Liu2017SLMC,Liu2016SLMCFermion,Nagai2017SLMCCT,Shen2018SLMCDNN,Xu:2017vug,Chen:2018aib,Nagai:2020ugq,nagai2020self,pawlowski2020reducing,foreman2021hmc}, where the goal is to directly learn the kernel $K(x'|x)$; learned sequential Monte Carlo sampling~\cite{arbel2021annealed,CRAFT2022,caselle2022stochastic,wu2020stochastic}, which combines deterministic flows with sequential Monte Carlo annealing techniques; Monte Carlo variance minimization~\cite{muller2019neural,veach1995optimally,muller2020neural}; and stochastic normalizing flows~\cite{wu2020stochastic}, which interleave deterministic transforms with latent-variable VAE-like components and Markov chain Monte Carlo transforms.
HMC is a particularly successful family of kernels and can also be combined with learnt components. In the presence of pseudo-fermions, simulating the Hamiltonian dynamics for HMC requires an expensive computation of `force terms', which is another area where acceleration may be possible using ML~\cite{Li2018,Li:2020,tomiya2021gauge,tanaka2017towards,nagai2020self}.

\subsection{Incorporating manifold constraints and gauge symmetry in ML models}
When the target density for sampling features an exact or approximate symmetry, breaking that symmetry in a sampling algorithm will result in computational inefficiencies. In particular, continuous exact symmetries naturally reduce the effective dimensionality of the target distribution, such that incorporating them directly reduces the difficulty of modelling that target. In addition, training a model with the symmetry explicitly encoded modifies the structure of the loss landscape in a way that may make training feasible in cases where attempting to approximately `learn' the symmetry may not be.
Moreover, if guarantees of exactness are required, any remaining symmetry-breaking after training will result in additional costs to correct that breaking via the approaches discussed in the previous subsection.
It is thus often advantageous, and in some cases critical, to incorporate symmetries into ML architectures and/or training approaches.

Modern ML offers many techniques that seek to take advantage of known symmetries. The simplest such method is data augmentation\cite{shorten2019survey}, where the available training set is `augmented' with randomly transformed input/output pairs. Another technique consists of explicitly adding a term to the optimization target that encourages equivariance or invariance with respect to a group of transformations\cite{mitrovic2020representation,rezende2019equivariant}. 
Both approaches only serve to assist the training, and symmetries still have to be learnt by the model architecture.
Alternatively, it is also possible to construct architectures such that they respect known symmetries by construction; standard convolutional neural networks, for example, are equivariant maps with respect to translations. The most commonly studied symmetries are finite groups\cite{cohen2016group} and SE(3), the group of isometries of 3D Euclidean space\cite{fuchs2020se,du2022se}.
Naturally, the larger the symmetry group, the harder it is to learn the symmetries, either via data augmentation or via additional optimization targets. In the case of lattice QCD, the gauge symmetry has a prohibitively large dimension scaling with the number of lattice sites, hence it is likely essential to build gauge-equivariant and invariant neural networks for this application. For example, with current architectures and training approaches it has been demonstrated that it is essential to exactly incorporate gauge symmetry~\cite{Kanwar:2020xzo}, but not translations and hybercubic transformations~\cite{Boyda:2020hsi}, for successful training of flow-based sampling algorithms for lattice field theory.

\begin{figure}[t]
	\centering
	\includegraphics[width=0.95\linewidth]{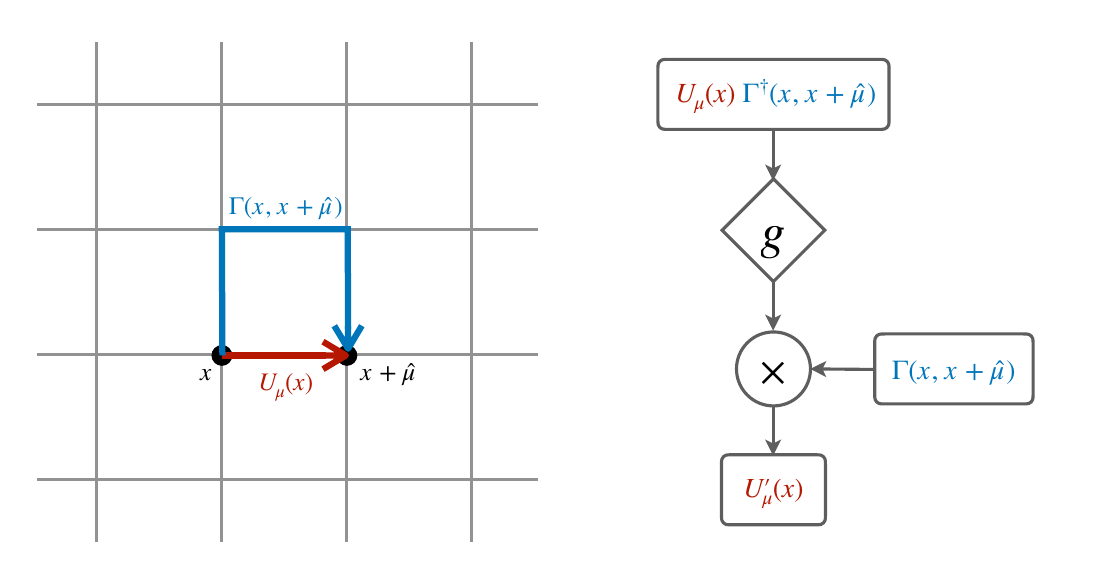}
	\caption{\textbf{Illustration of a gauge-equivariant transformation layer \cite{Boyda:2020hsi,Kanwar:2020xzo}.} Here, a parametric gauge-equivariant diffeomorphism is constructed from a diffeomorphism $g$ satisfying $g(\Omega U \Omega^{\dagger})=\Omega g(U) \Omega^{\dagger}$ (matrix-conjugation equivariance). To update the matrix-valued link $U_\mu(x)$ in red, this transformation first updates the plaquette $U_\mu(x) \Gamma^\dag(x, x+\hat{\mu})$ containing that link, before `pushing' that update onto the link by assigning $U'_\mu(x) = g(U_\mu(x) \Gamma^\dag(x, x+\hat{\mu})) \Gamma(x, x+\hat{\mu})$. The output link $U'_\mu(x)$ transforms appropriately under gauge transformations when $U_\mu(x)$ and $\Gamma(x, x+\hat{\mu})$ are transformed. See the main text for further details.}
	\label{fig:equiv}
\end{figure}

Building generative models that exactly incorporate the symmetry constraints of lattice QCD is a non-trivial task that has required the introduction of several new ML models to treat both gauge and pseudo-fermionic degrees of freedom\cite{jin2022neural,Boyda:2020hsi,Kanwar:2020xzo,katsman2021equivariant,finkenrath2022tackling,de2021scaling,Albergo:2021bna,hackett2021flow,albergo2019flow,vaitl2022}. This approach relies on the observation that starting from a base distribution that is gauge-invariant (such as the Haar measure on $\text{SU}(N)$) and applying a gauge-equivariant diffeomorphism to this base density yields a new density that is also gauge-invariant\cite{papamakarios2021normalizing,kohler2020equivariant,rezende2019equivariant,katsman2021equivariant}.
One can reduce the problem of building gauge-equivariant diffeomorphisms on the gauge degrees of freedom situated on the edges of a lattice to the problem of building matrix-conjugation-equivariant diffeomorphisms\cite{Boyda:2020hsi,Kanwar:2020xzo} on SU(3), which is a simpler problem. 
As described in equation~\eqref{eq:gauge-transform}, a gauge symmetry transformation $T_{\Omega}$ is parameterized by a field of SU(3) variables $\Omega(x)$ and acts on an edge, or `link', variable as
$T_\Omega U_\mu(x) = \Omega(x) U_\mu(x) \Omega^{\dagger}(x + \hat{\mu})$.
As a consequence, a product of link variables along a closed loop $\Lambda(x)$ starting and ending at a point $x$ transforms as $\Lambda(x) \rightarrow \Omega(x) \Lambda(x) \Omega^{\dagger}(x)$. Mathematically this is referred to as a matrix-conjugation, or adjoint transformation.
If $U_\mu(x)$ is a link, $\Gamma(x, x + \hat{\mu})$ is a product of links along an open path from $x$ to $x + \hat{\mu}$ that does not contain $U_\mu(x)$, and $g:\text{SU(3)}\rightarrow \text{SU(3)}$ is a conjugation-equivariant diffeomorphism, then a gauge-equivariant diffeomorphism $f$ can be constructed as
\begin{equation}
    f(U_\mu(x)) = g(U_\mu(x)\Gamma^{\dagger}(x, x + \hat{\mu}))\Gamma(x, x + \hat{\mu}),
\end{equation}
as shown in Fig.~\ref{fig:equiv}. The diffeomorphism  $f$ then manifestly satisfies gauge equivariance, $f(T_\Omega U) = T_\Omega f(U)$.
Building gauge-equivariant diffeomorphisms for pseudo-fermion degrees of freedom is also possible using extensions of this approach based on parallel transport\cite{Abbott:2022zhs}.

\subsection{A roadmap for ML-based sampling in lattice QCD}

\begin{figure*}[t!]
	\centering
	\vspace*{-0.2cm}
    \includegraphics[width=0.75\textwidth]{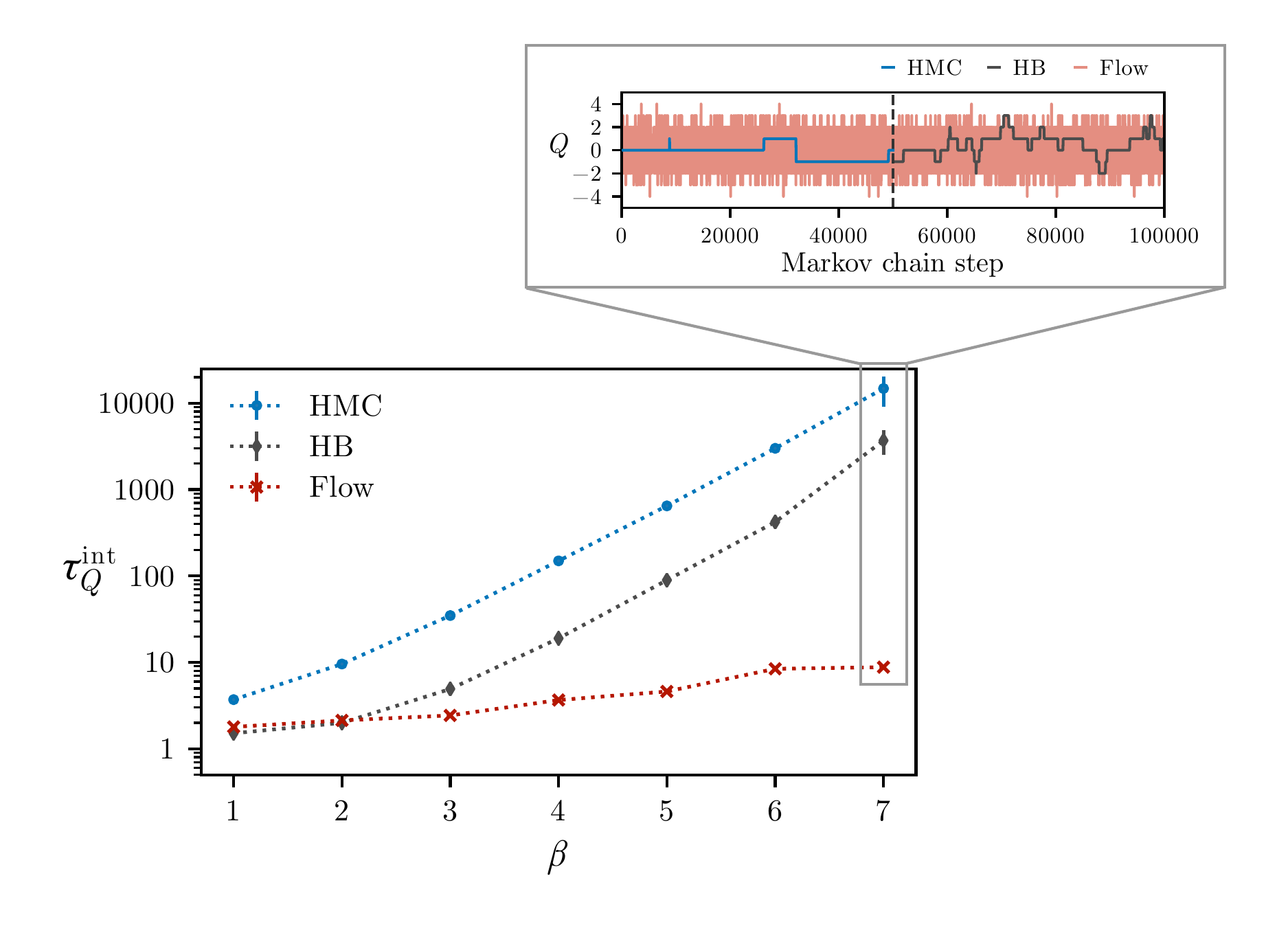}
    \vspace*{-0.5cm}
	\caption{\textbf{Demonstration of the advantages of flow-based sampling in a U(1) lattice gauge theory in two spacetime dimensions~\cite{Kanwar:2020xzo}.} The inset shows the rapid mixing of topological charge $Q$ when sampling with normalizing flows, compared with Hamiltonian/hybrid Monte Carlo (HMC) and Heat Bath (HB) algorithms for the action defined by $\beta = 7$ (see equation~\eqref{eq.pure_gauge_action} for the definition of the analogous parameter in quantum chromodynamics). The main graph shows the asymptotically improved scaling of $\tau^{\mathrm{int}}_Q$ towards the continuum limit $\beta \rightarrow \infty$, where $\tau^{\mathrm{int}}_Q$ is the `integrated autocorrelation time' of the topological charge, which is a measure of cost in Markov-process sampling and can be interpreted here as a metric for critical slowing-down. Reproduced with permission from ref.~\citen{Kanwar:2020xzo}, APS.
	}
	\label{fig.u1_topological_mixing}
\end{figure*}

The great potential of ML-based sampling for lattice field theories has inspired rapid developments that have already demonstrated profound successes. Following early applications of flow-based sampling to field theories other than QCD~\cite{albergo2019flow,Kanwar:2020xzo}, the approach was developed for theories in two spacetime dimensions, specifically for SU(3) gauge fields without dynamical quark (fermionic) degrees of freedom~\cite{Boyda:2020hsi} and for U(1) gauge fields with dynamical fermions~\cite{Albergo:2021bna}. Combining these advances enabled the first application of flow models to sampling QCD in 4D~\cite{Abbott:2022hkm}, albeit with small space-time volumes. ML-accelerated updating schemes have been developed, again for small volumes and with the SU(2) gauge group instead of SU(3) (ref.~\citen{nagai2020self}), and continuous-time models inspired by previous work in the lattice field theory community~\cite{luscher2010trivializing,luscher2011perturbative} have been applied to simple lattice field theories~\cite{Gerdes:2022eve} and both U(1) gauge theory ~\cite{jin2022neural} and SU(3) gauge theory~\cite{Bacchio:2022vje} in 2D. These approaches have had astounding success; Fig.~\ref{fig.u1_topological_mixing} illustrates the advantages of flow-based sampling in one particular toy theory, but the conclusion that ML-accelerated sampling schemes can overcome the critical slowing-down and topological freezing challenges faced by HMC has been clear and universal. It is important to emphasize that this success includes theories with fermionic degrees of freedom, where ML-accelerated sampling schemes have been developed to integrate with the usual approach of pseudo-fermions~\cite{Abbott:2022zhs,Albergo:2022qfi,Abbott:2022hkm}.
However, the crucial aspect missing in all applications so far is a demonstration of the effectiveness of ML techniques at the scale of state-of-the-art lattice QCD calculations in nuclear and particle physics.

\begin{figure}
	\centering
	\includegraphics[width=0.95\linewidth]{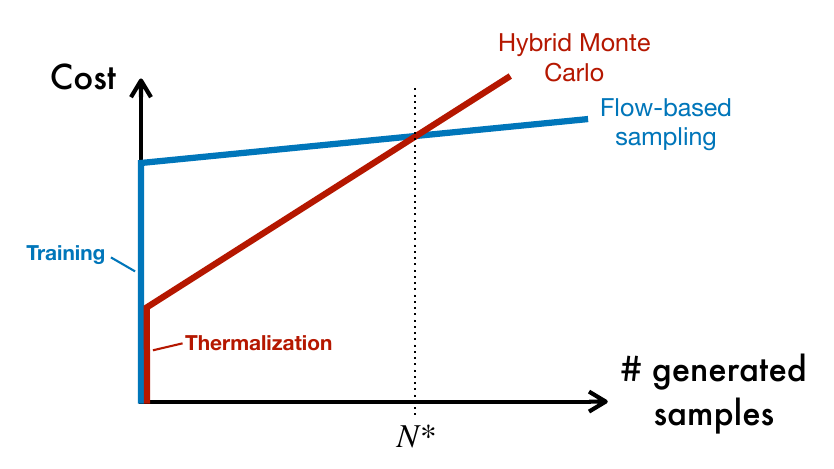}
	\caption{\textbf{Sketch of the upfront and sampling costs of hybrid Monte Carlo compared with flow-based models.} Hybrid Monte Carlo (HMC) shown in blue; flow-based sampling in red. Whereas machine-learning models require a potentially large upfront computational cost for training, they offer the hope of efficient sampling, such that this training cost can be amortized. That is, for sufficiently many samples, flow-based sampling that is more efficient than HMC (for example, as in Fig.~\ref{fig.u1_topological_mixing}) can outperform HMC in terms of total cost. The minimum number of samples for which flow-based sampling more efficient than HMC is denoted by $N^*$.
	}
	\label{fig:flow-advantage}
\end{figure}

We expect that not only will we soon see applications of ML-accelerated sampling to lattice field theory at scale, but that running at scale is a key ingredient necessary to realize the full potential of ML in this context. In particular, we anticipate that the first impact of ML for this application will be that, once the potentially high cost of training is paid, ML-based sampling will be orders of magnitudes faster than traditional HMC, mitigating critical slowing-down, overcoming topological freezing, and opening the door to a sampling regime where this training cost can be efficiently amortized, as depicted in Fig.~\ref{fig:flow-advantage}. At precisely which scale this advantage will be reached is not yet clear; the computational cost of training ML models in this context may vary by orders of magnitude between different architectures and training approaches~\cite{Abbott:2022zsh}. As the optimal approach to model parameterization and training can depend sensitively on the number of samples which are ultimately required, the balance of training and sampling costs is highly problem-dependent, and the regime in which flow-based sampling outperforms HMC for lattice QCD applications will depend on precisely how the flow models are used (and reused).
It is already evident, however, that achieving this paradigm of efficient ML-accelerated sampling will require considerable investment; it is clear that in the field of generative ML as a whole, the substantial progress in text~\cite{hoffmann2022training,kaplan2020scaling,brown2020language,lieber2021jurassic,rae2021scaling,smith2022using,thoppilan2022lamda} and image modelling~\cite{oord2016wavenet, dhariwal2021diffusion,saharia2022photorealistic,child2020very} has required pushing the boundaries of model size. So far, the generative ML experiments for lattice field theory are of comparatively small scale, despite the target scale of the problem itself being comparable to, or even larger than, applications in these domains. Success will thus likely require model scales, and corresponding investments in upfront training, that will constitute a change in paradigm of computational resource use for the theoretical physics community. Nevertheless, as a fundamentally structured problem, we anticipate that scaling the custom ML solutions developed for lattice field theory to large models will pay dividends and that lattice QCD will join the list of scientific problems which have seen significant impacts from ML at state-of-the-art scale, such as low-dimension Bayesian parameter inference for astrophysics~\cite{gabbard2022bayesian}, quantum Monte Carlo~\cite{musaelian2023learning} and protein-folding~\cite{jumper2021highly}.

Beyond the anticipated impacts of mitigating sampling challenges such as critical slowing-down and topological freezing, ML models have the potential to catalyse other paradigm shifts in lattice field theory. For example, they naturally offer new opportunities for community resource sharing. Ensembles of lattice field configurations are large enough in size (petabytes for state-of-the-art ensembles) that they can not be easily shared, and massive investments in tape resources are made to store them. In contrast, even the largest ML models only contain a few terabytes of parameters. These can easily be shared, allowing research groups around the globe to efficiently generate their own configurations or reproduce ensembles from a known seed, in both cases capitalizing on community-owned pre-trained models. Another important opportunity is that ML-based samplers can be conditioned on various parameters of the theory, from the lattice spacing and volume to physical parameters such as the strength of coupling of the fundamental particles of the theory. The potential to generate `correlated' sets of samples at different parameters, interpolated, or even extrapolated\cite{Singha:2023cql,lehner2023gauge}, from the parameters used during training, is qualitatively distinct from what is possible using traditional sampling algorithms such as HMC (in principle this is also possible for pre-selected parameter sets, using approaches such as parallel tempering), and offers new parameter extrapolation methods. Ultimately, one could even imagine more general ways of conditioning these models (for example, on a symbolic description of the target action), enabling new approaches such as direct measurements of the effect of modifying the action on physical observables. As such, ML techniques hold the promise of redefining conventional wisdom in this field, with implications that have yet to be fully explored.

\section{Outlook}\label{sec:outlook}

As ML continues to evolve in scope and complexity, its applications in science are being driven into two broad categories: those that can adapt existing ML technologies (usually developed to model images, sound and text) and those that demand ground-up development and inspire innovation. Lattice field theory is becoming established as a prime example of the latter, being simultaneously an important science application in which algorithmic acceleration will have wide-reaching implications for fundamental physics and a massive computational challenge of a scope and scale that has driven advances in computation and algorithms for decades. 

In particular, the challenge of sampling lattice field configurations for nuclear and particle physics calculations is a definitive proving-ground for generative ML models in science. With strict requirements of asymptotic exactness, and an ultimate scale at which each sample is several terabytes in size, it is clear that symmetries, structure and domain knowledge must be incorporated into ML architectures designed for this task. In that regard, the application of ML to sampling in lattice field theory is a key exemplar informing the debate around the value of engineering and incorporating domain or expert knowledge that is currently underway in the ML community: whereas ardent supporters of modern deep learning often argue against the long-term value of incorporating such knowledge
(the `bitter lesson' theory~\cite{sutton2019bitter}, advocated by Richard Sutton), in the physical sciences, and in particular in theoretical physics calculations, there are often precise mathematical formulations of domain knowledge that not only make little sense to ignore, but are intractable to learn from data.
Adding to the complexity of this engineering and design challenge, existing algorithmic benchmarks for sampling in lattice field theory are extremely high; new ML samplers must compete against well-established algorithms that have been optimized in co-design with high-performance computing systems for more than four decades.

As such, ML for lattice field theory is not only a benchmark for the application of ML in science, but it is also an endeavour with paradigm-shifting potential for physics. If the success already achieved in sampling field theories at toy scales can be mimicked at state-of-the-art scales, it will transform the computational landscape of a field that is one of the largest consumers of open-science supercomputing (computing available to public scientific applications) worldwide, with impacts across particle, nuclear and condensed matter physics and beyond. On the ML side, it will be a flagship example of the power of sophisticated domain-specific customization and engineering to achieve transformative impact in computational science.

\section*{Acknowledgements}
We thank W.~Detmold and R.~D.~Young for comments on the manuscript.
P.E.S.\ was supported in part by the U.S. Department of Energy, Office of Science, Office of Nuclear Physics, under grant Contract Number DE-SC0011090 and by Early Career Award DE-SC0021006, by a NEC research award, and by the Carl G.\ and Shirley Sontheimer Research Fund.
G.K.\ was supported by funding from the Schweizerischer Nationalfonds
(grant agreement no.\ 200020\_200424).

\section*{Author contributions}
The authors contributed equally to all aspects of the article.

\bibliography{references}

\end{document}